\documentclass[aps,prl,preprint,unsortedaddress,showpacs,floatfix,nofootinbib]{revtex4-1}
\usepackage[colorlinks=true,breaklinks=false]{hyperref}

\usepackage{graphicx}
\usepackage{amsmath,amstext,amssymb}
\usepackage{bm}
\usepackage{xcolor}

\begin{document}

\title{Dispersive wave emission from dark solitons in microresonator-based Kerr frequency combs}

\author{Shaofei Wang}
\email{Corresponding author: beileye288@shu.edu.cn}
\author{Xianglong Zeng}

\affiliation{Key Laboratory of Specialty Fiber Optics and Optical Access Network, Shanghai University, 200072 Shanghai, China}

\begin{abstract}
Taking advantage of an extended Lugiato--Lefever equation with third-order dispersion, we numerically show that dark cavity solitons formed in normal dispersion of microresonators are capable of emitting dispersive waves in both normal and anomalous dispersion regions, resembling the behavior of the commonly encountered bright cavity solitons. The generated dispersive waves can be accurately predicted by the dissipative radiation theory. In addition, we demonstrate the stability enhancement of Kerr frequency combs in normal dispersion regime in case the dispersive wave is emitted by dark solitons in presence of third-order dispersion.
\end{abstract}

\pacs{42.65.Sf, 42.65.Tg, 42.65.Hw}

\maketitle

\noindent Kerr frequency combs generated in high-Q whispering gallery mode microresonators thanks to the cascaded four-wave mixing (FWM) effect have attracted substantial research interests over the past few years \cite{Kippenberg2011}. In particular, temporal cavity solitons (TCSs) were demonstrated both theoretically \cite{Matsko2011,Coen20131,Chembo2013} and experimentally \cite{Herr20141,Saha2013} in microresonators, pushed one step further for research of versatile Kerr frequency combs. TCSs are localized dissipative pulses which are able to excite phase-locked, high-coherent Kerr frequency combs \cite{Lamont2013,Herr20142,Pel'Haye2015}. In fact, cavity solitons were first studied in spatial domain and passive fiber cavity structures \cite{Haelterman19921,Haelterman19922,Leo2010}, taking advantage of the well-known mean-field Lugiato-Lefever (LL) equation \cite{Lugiato1987}. The LL equation indeed builds up a direct bridge from earlier research of cavity solitons towards current Kerr frequency combs, offering an unprecedented approach to explore more details in both time- and frequency domains of Kerr frequency combs \cite{Godey2014}. Therefore, considerable efforts have been dedicated recently to investigating microresonator-based TCS dynamics in presence of various perturbations, among which the effect of high-order dispersions (HODs) is an important aspect. Reminiscent of behaviors in conservative fiber geometries \cite{Dudley2006}, HODs enable TCSs formed in microresonators to emit dispersive waves (DWs) as well \cite{Milian2014,Brasch2014,Jang2014,Wang2014,Parra-Rivas20141,Milian2015}. Typically, TCSs exist in the anomalous group velocity dispersion (GVD) region in microresonators \cite{Herr20141}. However, recently it was found remarkably that mode-locking of Kerr frequency combs can also be realized in normal GVD regime \cite{Matsko20121,Coillet2013,Liang2014,Xue2015,Huang2015}, namely the so-called dark TCSs in temporal domain. Even in some extreme cases, bright (dark) TCSs can be supported in normal (anomalous) GVD as well \cite{Tlidi2010,Tlidi2013,Huang2015}. Dark TCSs are being drawn grown attention in frame of Kerr frequency combs \cite{Liu2014,Lobanov2015}. Yet, unlike bright TCS counterparts, studies of dark TCSs in normal GVD region in presence of perturbations are rather sporadic. Motivated by earlier results about the influence of the third-order dispersion (TOD) on solitons in nonlinear fibers \cite{Afanasjev1996,Milian2009}, we present some specific cases to illustrate the influence of the TOD term on dark microresonators-based TCSs in normal GVD regime, by meas of the LL equation including the TOD term. We find that dark TCSs in normal GVD region are able to completely exhibit the features of bright TCSs in anomalous GVD region, including the DW generations as well as comb stabilizations. Due to the periodic boundary nature of resonators, the generated DWs could be located in both anomalous and normal GVD regimes.

The mean-field LL equation is a damped, driven Schr\"odinger equation which is widely utilized to describe Kerr frequency comb generation in microresonators \cite{Matsko2011,Coen20131,Chembo2013}. The normalized LL equation with the TOD term reads

\begin{eqnarray}
\frac{{\partial E\left( {t,\tau } \right)}}{{\partial t}} =  - \left( {1 + i\Delta } \right)E + i{\left| E \right|^2}E - i\frac{{{\beta _2}}}{2}\frac{{{\partial ^2}E}}{{\partial {\tau ^2}}} + \frac{{{\beta _3}}}{6}\frac{{{\partial ^3}E}}{{\partial {\tau ^3}}} + F,
\label{eq1}
\end{eqnarray}
where $E$ is the intracavity field, $\tau$ represents the time within a single roundtrip, termed the \textit{fast time}, $t$ denotes the time over successive roundtrip numbers, which is called the \textit{slow time}. $\Delta$ is the phase detuning with respect to the continuous wave (cw) pump frequency. $F$ is the external cw driven field. $\beta_2$ and $\beta_3$ are GVD and TOD terms, respectively.

Normally, the HODs become crucial and cannot be ignored when an octave-spanning Kerr frequency comb is excited in a microresonator with a low GVD value over a wide frequency range \cite{Del'Haye2011,Okawachi2011,Brasch2014}, in which the waveguide dispersion is able to reshape the overall dispersion profile, giving rise to a zero dispersion point (ZDPs) at short frequency. Alternatively, when the pump cw frequency is in the vicinity of the ZDP, the role of the HODs are significant thus the DW emission is possible \cite{Grudinin2013}. For clarity, here we consider the dominated TOD term in the LL equation. According to the dissipative radiation theory proposed in Refs. \cite{Milian2014,Malaguti2014}, it is easy to obtain the following relation for the complex frequency $\Omega$ of the generated DW:

\begin{equation}
V\Omega = \frac{{{\beta _3}}}{6}{{\Omega}^3} + i \pm \sqrt {{{\left( {2{P_0} - \Delta  + \frac{{{\beta _2}}}{2}{{\Omega}^2}} \right)}^2} - P_0^2},
\label{eq2}
\end{equation}
where $V$ is the soliton temporal drift velocity within a single roundtrip due to the TOD term \cite{Wang2014,Parra-Rivas20141}. $P_0$ is the TCS background power (corresponds to the cw power) rather than the peak power \cite{Milian2014,Jang2014}. For a dark cavity TCS, $P_0$ corresponds to the top flat power value. Further considering the real part solutions of Eq. \ref{eq2}, we can acquire the final expression for the resonant relation, which is referred as $R\left(\Omega\right)$, namely,

\begin{equation}
R\left( \Omega \right) = \frac{{{\beta _3}}}{6}{{\Omega}^3} - VQ \pm \sqrt {{{\left( {2{P_0} - \Delta  + \frac{{{\beta _2}}}{2}{{\Omega}^2}} \right)}^2} - P_0^2}.
\label{eq3}
\end{equation}

Obviously, the possible roots for $R\left( \Omega \right)= 0$ are the frequency locations of the generated DWs. Equation \ref{eq3} is an odd function with respect to the center frequency $\Omega = 0$. For this reason, it differs from the classical resonant radiation in fibers since it holds at least a pair of symmetrical DWs in case of a single ZDP \cite{Milian2014,Jang2014}.

\begin{figure}[htbp]
\centerline{\includegraphics[width = 1.0\columnwidth]{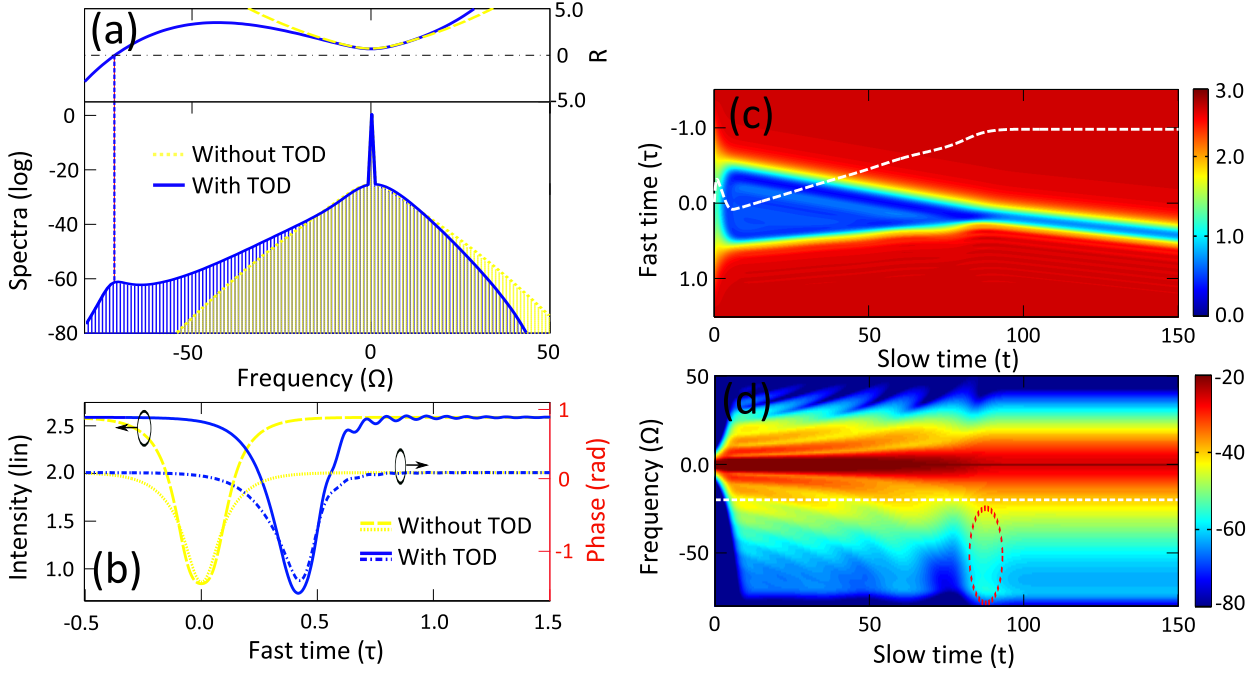}}
\caption{(Color online) (a) Simulation results of final comb spectra with (blue solid line) and without (yellow dashed line) the TOD term. he blue solid and yellow dashed lines in the top are the '+' branch of $R\left( \Omega \right)$ with and without TOD, respectively. (b) The final temporal profiles with (blue solid line) and without (yellow dashed line) the TOD term, as well as the phase profiles in presence of (blue dash-dot line) and in absence of (yellow dotted line) the TOD term. (c) and (d) are the evolution dynamics of the temporal and spectral domains, respectively. The white dashed line in (c) is the intracavity power along propagation with arbitrary unit. The white dashed line in (d) denotes the ZDP, and the red oval area highlights the DW emission.}
\label{fig1}
\end{figure}

Now we begin to address the first case. Specifically, we employ the parameters $\Delta = 2.5$, $F^2 = 2.6$, which are confined within normal-dispersion TCS area according to the nonlinear bifurcation diagram of the LL equation, see Ref. \cite{Godey2014} for details. In fact, dark TCSs in normal GVD region are more difficult to be aroused since their parameter space are fairly narrow compared with that of the bright TCSs. Considering the fact that TCSs cannot be built directly from initial noise since the modulation instability area does not connect with the TCS region in normal dispersion regime, we therefore utilize an exponential-shape gap pulse instead of the random noise as the initial pump state, following commonly employed approaches, for instance Refs. \cite{Godey2014,Coillet2013}, namely,

\begin{equation}
{E_0}\left( {0,\tau } \right) = a - b\exp \left[ { - {{\left( {{\tau  \mathord{\left/ {\vphantom {\tau  c}} \right. \kern-\nulldelimiterspace} c}} \right)}^2}} \right],
\label{eq4}
\end{equation}
where we employ $a = 1.7$, $b = 1.0$, and $c = 0.9$ for the first case. Note that the dispersion scaling is not important to excite a resonator \cite{Coen20132}, therefore we use the relative small dispersions, i.e., $\beta_2 = 0.005$, $\beta_3 = 0.05\beta_2$, to arise a broad spectrum enabling a considerable DW emission. The dispersion parameters yield a ZDP $\Omega_z = -20$.

Simulations of dimensionless Eq. \ref{eq1} are performed employing the above parameters. The \textit{fast time} window is chosen much larger than the initial pulse duration, ensuring that boundary conditions do not influence the TCSs. The \textit{slow time} step size is set as $0.001$ in simulations, which is small enough to avoid any possible artifacts. Simulation results are depicted in Fig. \ref{fig1}. Therein, the output spectrum in presence of TOD (i.e. blue solid line in the bottom of Fig. \ref{fig1}(a)) clearly suggests that a DW is emitted around $\Omega = -71$, which is in excellent agreement with the corresponding $R\left( \Omega \right)$ (i.e. the blue solid line in the top of Fig. \ref{fig1}(a)). In contrast, without the TOD term, the final steady comb exhibits a completely symmetric spectrum profile without any DW emission (i.e. yellow dashed line in the bottom of Fig. \ref{fig1}(a)), as expected from the corresponding resonate relation (i.e. the yellow dashed line in the top of Fig. \ref{fig1}(a)). Although the theoretical analysis of Eq. \ref{eq3} reveals that double symmetric DWs should be created in this case, we find in simulations the other DW is too weak to be observed, so we merely present the '+' branch of the $R\left(\Omega\right)$, corresponding to the powerful DW observed in Fig. \ref{fig1}. We will illustrate this issue in the following case.

\begin{figure}[htbp]
\centerline{\includegraphics[width = 1.0\columnwidth]{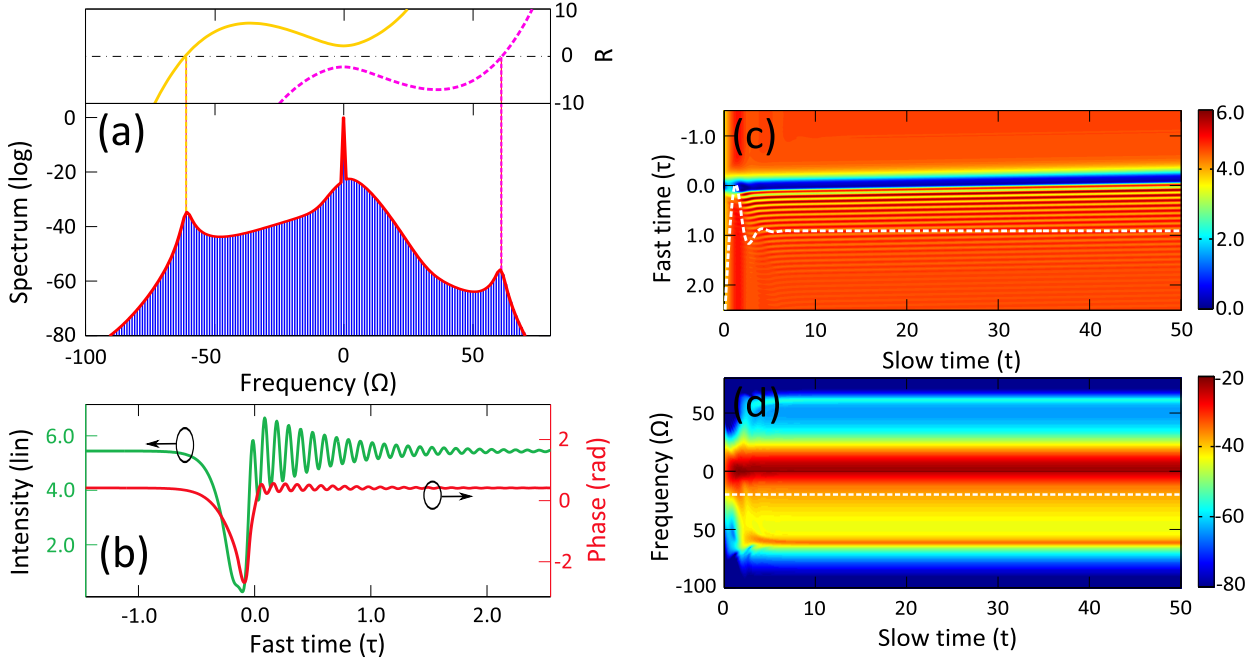}}
\caption{(Color online) Same description as Fig. \ref{fig1} in case of double DW emissions in the presence of the TOD term. The yellow solid and pink dashed lines in the top are the corresponding '+' and '-' branches of the resonate relation.}
\label{fig2}
\end{figure}

On the other hand, from the final dark TCSs formed in temporal domain, as shown in Fig. \ref{fig1}(b), one can easily identify that a standard dark TCS is formed in the center of the \textit{fast time} in absence of the TOD term, which is plotted as the yellow dotted line in Fig. \ref{fig1}(b). Nevertheless, the presence of the TOD gives rise to the DW emission associating with the considerable time drift as well as the soliton tail oscillation, which are exactly the same behaviors for bright TCSs \cite{Parra-Rivas20141} and even for solitons in classical fibers \cite{Afanasjev1996}. In this regard, they share the same nature due to the fact that they are all governed by the master Schr\"odinger equation. Interestingly, the final phase profiles of the corresponding TCSs given in Fig. \ref{fig1}(b) show the corresponding TCS profiles. Further focusing on the temporal and spectral evolutions along \textit{fast time}, as shown in Figs. \ref{fig1}(c) and \ref{fig1}(d), we find the dark TCS is formed from $t = 96$, which is estimated by detecting the intracavity power (i.e. white dashed line in Fig. \ref{fig1}(c)). The fine fringes in the vicinity of the soliton edge denotes the DW emission process, which is highlighted more intuitively in spectral domain, as seen in Fig. \ref{fig1}(d). In addition, we find the DW-like emission comes into being from $t = 10$, far priors to the final TCS formation, implying that TOD disturbs the field $E$ since the beginning of the propagation.

We now proceed to validate the DW emission from dark TCSs in microresonators. In the second case, we change the dissipation parameters $\Delta = 5.0$, $F^2 = 6.5$, and we increase the dispersion values $\beta_2 = 0.0125$, $\beta_3 = 0.06\beta_2$ to further boost the DW emission efficiency. The generated ZDP is $\Omega_z = -16.67$. Reasonably, the corresponding initial pump condition should be modified as well, so we set $a = 2.0$, $b = 1.5$, and $c = 0.1$ for Eq. \ref{eq4}. Simulation results are given in Fig. \ref{fig2}, clearly showing a pair of DWs locate at $\Omega = \pm 61$, respectively. The generated DWs are accurately predicted by Eq. \ref{eq3} for both branches. The left-hand DW is, however, more efficient than the other, which accounts for the single DW emission in Fig. \ref{fig1} and one earlier experiment of bright TCSs \cite{Jang2014}. Moreover, we find that the final TCS is accompanied with a negligible temporal drift (i.e. offset from $\tau = 0$) but a strong oscillate tail (i.e. a strongly stretched but localized branch near the main dark TCSs), suggesting that powerful DW emissions are achievable. The detailed evolution dynamics along successive roundtrips of the TCS are shown in Figs. \ref{fig2}(c) and \ref{fig2}(d), in which a pair of DWs are emitted around $t = 2$ while the TCS is formed from $t = 5$. Although the double DWs are symmetrically located in either side of $\Omega = 0$, they share the identical side in temporal domain, as can be identified from Fig. \ref{fig2}(b). We attribute this feature by the periodic boundary conditions imposed by the resonator nature.

\begin{figure}[htbp]
\centerline{\includegraphics[width = 0.6\columnwidth]{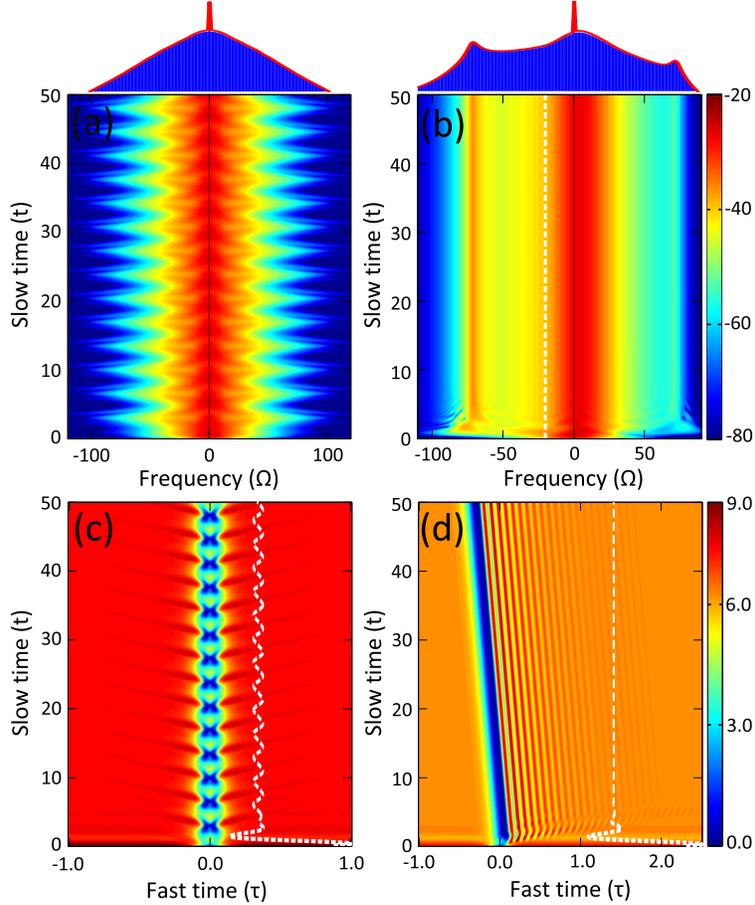}}
\caption{Kerr frequency comb stability analysis without (left panel) and with (right panel) the TOD term. The top panel are the corresponding final comb profiles. (a)-(b) and (c)-(d) are, respectively, the spectral and temporal evolution dynamics. The white dash line in (b) is the ZDP. The white dashed lines in (c) and (d) are the intracavity power along propagation. Note in this case, $a = 2.7$, $b = 1.8$, and $c = 0.1$, $\beta_2 = 0.005$, $\beta_3 = 0.06\beta_2$.}
\label{fig3}
\end{figure}

It has been demonstrated that bright TCSs in anomalous GVD region is able to stabilize Kerr frequency combs in presence of TOD \cite{Parra-Rivas20141}. Next we will investigate this issue in terms of dark TCSs in normal dispersion. To do this, we increase the dissipation values, i.e., $\Delta = 6.0$, $F^2 = 8.5$, enabling the so-called \textit{breather} solitons exist in the resonator. In fact, the \textit{breather} is a special type of TCSs, existing above the Hopf threshold in the soliton branch of bifurcation diagrams \cite{Matsko20122,Leo2013}. In the context of \textit{breather} TCSs, the spectrum permanently oscillates along propagation, for instance our case shown in Figs. \ref{fig3}(a) and \ref{fig3}(c) (see the white dashed line) in absence of the TOD term. The same behavior is available for bright TCSs in case of anomalous GVD \cite{Matsko20122,Leo2013}. Obviously, the \textit{breather} TCSs would degrade the coherence of the final comb profile \cite{Matsko20122,Erkintalo2014}, however, if we incorporate the TOD term in Eq. \ref{eq1}, one can see clearly that the \textit{breather} TCS vanishes, followed by a stable TCS with a constant power along propagation, as shown in Fig. \ref{fig3}(d). In spectral domain, as depicted in Fig. \ref{fig3}(b), the periodic comb feature is also replaced by a constant spectrum accompanying with the double-band DW emissions. Theoretically speaking, the TOD is able to stabilize Kerr frequency combs since it modifies the Hopf threshold and shrinks the existence area of the \textit{breathers}, therefore giving rise to a tremendous parameter space for stable TCSs\cite{Parra-Rivas20142}. In other words, the DW emissions from dark TCSs in normal dispersion regime are capable of stabilizing Kerr frequency combs in normal GVD regime.

In conclusion, we have demonstrated numerically the DW emission from dark TCSs in normal GVD of microresonators by incorporating the TOD term in an extended LL equation. We show that the generated DWs can be located in both normal and anomalous GVD regimes in case of a single ZDP, as expected from the dissipative radiation theory. Furthermore, stabilizing Kerr frequency combs in normal GVD regime are confirmed by involving the DW emission from dark TCSs. The result obtained here based on the simplified model will benefit for future exploring more realistic scenarios of Kerr frequency combs in normal dispersion regime.

\end{document}